\documentclass[pra,twocolumn,showpacs,preprintnumbers,superscriptaddress]
{revtex4}

\usepackage{times}
\usepackage{bm}
\usepackage{float}
\usepackage{graphicx}
\usepackage{amsbsy}
\usepackage{amsmath}
\usepackage{amsfonts}
\usepackage{amsthm}
\begin{document}
	
\theoremstyle{plain}
\newtheorem{theorem}{Theorem}
\newtheorem{lemma}[theorem]{Lemma}
\newtheorem{corollary}[theorem]{Corollary}
\newtheorem{proposition}[theorem]{Proposition}
\newtheorem{conjecture}[theorem]{Conjecture}

\theoremstyle{definition}
\newtheorem{definition}[theorem]{Definition}

\theoremstyle{remark}
	\newtheorem*{remark}{Remark}
	\newtheorem{example}{Example}
	\title{Structural physical approximation of partial transposition makes possible to distinguish SLOCC inequivalent classes of three-qubit system}
	\author{Anu Kumari, Satyabrata Adhikari}
	\email{mkumari_phd2k18@dtu.ac.in, satyabrata@dtu.ac.in} \affiliation{Delhi Technological
		University, Delhi-110042, Delhi, India}
	
\begin{abstract}
Detection and classification of entanglement properties of a multi-qubit system is a topic of great interest. This topic has been studied extensively and thus we found different approaches for the detection and classification of multi-qubit entangled states. We have applied partial transposition operation on one of the qubit of the three-qubit system and then studied the entanglement properties of the three-qubit system, which is under investigation. Since the partial transposition operation is not a quantum operation so we have approximated partial transposition operation in such a way so that it represent a completely positive map. The approximated partial transposition operation is also known as structural physical approximation of partial transposition (SPA-PT). We have studied in detail the application of SPA-PT on a three qubit system and provided explicitly the matrix elements of the density matrix describing SPA-PT of a three qubit system. Moreover, we propose criterion to classify all possible stochastic local operations and classical communication(SLOCC) inequivalent classes of a pure as well as mixed three qubit state through SPA-PT map, which makes our criterion experimentally realizable. We have illustrated our criterion for detection and classification of three-qubit entangled states by considering few examples.

\end{abstract}
 \maketitle
	
\section{Introduction}
Quantum entanglement\cite{einstein} is a physical phenomenon in which the state of each particle in the group cannot be described independently of the state of the others, even when the particles are separated by a great distances. This phenomenon cannot be explained by classical physics due to the existence of non-local feature present in it. The non-local property exhibited by the entangled quantum system in $d_{1}\otimes d_{2}$ dimensional Hilbert space may be useful in various quantum information processing tasks such as quantum teleportation\cite{bennett}, remote state preparation\cite{pati}, enatnglement swapping\cite{pan}, secret sharing\cite{hillery} and quantum repeater\cite{li}.\\
We require entangled states to perform quantum information processing tasks in an efficient way, but the process of generation of entangled states is not an easy task. Even if we generate quantum state in an experiment, it is not known that whether the generated state is an entangled state or not? We can answer this question, if we are able to proceed little bit further toward the problem of "detection and classification of entangled states".
Detection and classification of entanglement has very vast literature and thus in this direction of research, one can find many proposed entanglement detection criterion such as computable cross norm or realignment criterion(CCNR criterion) \cite{rudolph,chen,chen2}, range criterion\cite{horodecki} etc. Entanglement in bipartite and multipartite system \cite{guhne} can also be detected through the construction of entanglement witness operator.
The separability criteria introduced by Peres\cite{peres} provides a very powerful criterion for the detection of entanglement. This criterion is also known as positive partial transpose (PPT) criterion. Later, Horodecki's \cite{mhorodecki} proved that this criterion is necessary and sufficient for $2\otimes 2$ and $2\otimes 3$ quantum systems. Since there exist bipartite entangled states in higher dimensional system and also entangled states in multipartite systems that does satisfy the PPT criterion so it remain only as a necessary criterion for higher dimensional bipartite as well as multipartite systems.\\
Although PPT criterion serve as necessary and sufficient condition in the detection of entanglement in $2\otimes2$ and $2\otimes3$ dimensional systems but it suffers from a serious drawback. The partial transposition map used in PPT criterion is a positive but not completely positive map. Thus, partial transposition operation cannot be implemented in a real experimental setup. Approximation of partial transposition map provide a possible solution to get rid of this problem. The map that approximate the partial transposition map is known as structural physical approximation of a partial transposition (SPA-PT) map. The concept of SPA-PT map has been introduced by Horodecki and Ekert \cite{phorodecki} for $d\otimes d$ dimensional systems. It is a completely positive map corresponds to a quantum channel that can
be experimentally implementable \cite{bae}. SPA-PT map has been constructed in such a way that it will become a completely positive map. Let us consider a $d\otimes d$ dimensional quantum system described by the density operator $\sigma_{12}$. If $\sigma_{12}^{T_{B}}$ ($T_{B}$ denotes partial transposition with respect to the second subsystem B) represent a partial transposed state then the SPA-PT of $\sigma_{12}^{T_{B}}$ is given by $\widetilde{\sigma_{12}}$. As a consequence of the application of SPA-PT map, the PPT criterion get modified and now it can be stated in the following way: if the minimum eigenvalue of $\widetilde{\sigma_{12}}$ is less than $\frac{d^{2}\lambda}{d^{4}\lambda+1}$ then the state $\sigma_{12}$ is entangled \cite{phorodecki}. Here $-\lambda$ denote the most negative eigenvalue obtained when the induced map $[(I\otimes I) \otimes (I \otimes T^{B})]$ is acting on the maximally entangled state in $d\otimes d$ dimensional system. In particular, for $2 \otimes 2$ dimensional system, the reduced SPA-PT criterion can be states as: if the minimum eigenvalue of $\widetilde{\sigma_{12}}$ is less than $\frac{2}{9}$ then the state $\sigma_{12}$ is entangled. The minimum eigenvalue can be estimated by the procedure given in \cite{keyl,tanaka}. SPA-PT method not only can detect entangled states but also has many applications in quantum information processing tasks such as in estimating the optimal singlet fraction\cite{adhikari}. Criterion based on the method of SPA-PT have been given for the detection of mixed bipartite entangled state in arbitrary dimension\cite{adhikari2}. Furthermore, SPA conjecture has been discussed in\cite{lewenstein1} and it has been proved that for any positive map($\gamma$) there exists an entanglement breaking channel($\phi$) such that SPA of $\gamma$ with the aid of $\phi$ is again an entanglement breaking channel. They have also defined a way for the construction of SPA-PT of positive map in continuous variable system. Further, in order to disprove SPA conjecture, Ha and Kye\cite{kye} proposed a decomposable entanglement witness operator whose SPA is entangled and argued that it is optimal. In\cite{lewenstein2} authors have shown both analytically and numerically that this entanglement witness is not optimal usind the method defined in \cite{lewenstein3}.

To show the existence of nonlocality experimentally in multipartite system is difficult. In this context, J. Tura et.al. \cite{lewenstein4} have shown that it is possible to detect nonlocality in multipartite systems using Bell inequalities
with only two-body correlators. In \cite{schmied}, 
authors have worked on detection of Bell correlations with trusted collective measurements through Bell correlation witnesses. These witnesses were then tested experimentally in many body systems such as Bose-Einstein condensate or thermal ensembles. The number of particles sharing genuinely nonlocal correlations in a multi-partite system has been studied in  \cite{lewenstein5} and characterize all Bell-like inequalities for a finite number of parties based on symmetric two-body correlations. Moreover authors have also provided witnesses that can be used in experiments to reveal a Bell correlation depth k $\leq$ 6 for any number of parties.\\

Acin.et.al have shown that the set of density matrices for three-qubits contains convex compact subsets of states belonging to the
separable, biseparable, W and GHZ classes, respectively. These classes are successively embedded into each other. All possible stochastic local operation and classical communication(SLOCC) inequivalent classes have been classified as one fully separable state, three biseparable states and two genuine entangled states\cite{datta}. The detection and classification problem for three-qubit system has been studied by the construction of entanglement witnesses\cite{acin,ryu,kye2,eltschka1} and through entanglement measures such as tangle\cite{eltschka}.
 Entanglement witnesses (EWs) constitute one of the most important entanglement detectors in quantum systems. Their complete characterization, in particular with respect to the notion of optimality, is still
 missing, even in the decomposable case. In \cite{lewenstein6}, authors have shown that for any qubit-qunit decomposable entanglement witness W, the three statements are equivalent: (i) the set
 of product vectors obeying $\langle e,f|W|e,f\rangle=0$ spans the corresponding Hilbert space, (ii) W is optimal, (iii) W = $Q^{T_B}$ with Q denotes a positive operator
 supported on a completely entangled subspace and $T_B$ denotes the
 partial transposition with respect to subsystem B. The characterization of entanglement depth  for many body system was studied in \cite{lucke,lewenstein8}. The device independent witnesses have been constructed to detect entanglement depth \cite{lewenstein7}.\\

Since the PPT criterion provides only a necessary condition for the detection of three-qubit entangled system so this may be a possible reason for not exploring it in detail. In this work, we will use partial transposition operation to investigate the problem of detection of a three-qubit entangled system. Then we will study the SPA-PT map for the three-qubit system to classify its different SLOCC inequivalent classes.\\
This paper is organized as follows: In Sec-II, we have revisited the partial transposition map and studied its effect in the detection of three-qubit $GHZ$ and $W$ class of states. In section-III, we introduce SPA-PT map for three-qubit system and calculated different entries of the matrix of SPA-PT map. In section-IV, we provide different criteria for the detection and classification of any three-qubit system. In section-V, we illustrate our criterion by examples.

\section{Studying the Effect of Partial Transposition Operation on one of the qubit of a Three-Qubit System}
Here, we will study the effect of partial transposition operation on any one of the qubit of a three-qubit system shared between Alice, Bob and Charlie. Let us assume that any three-qubit state is described by the density operator $\rho_{ABC}$. If the entries of the three-qubit state $\rho_{ABC}$ represented by the $2\times 2$ block matrices then it is given by
\begin{eqnarray}
\rho_{ABC}=
\begin{pmatrix}
A & B & C & D\\
B^{*} & E & F & G\\
C^{*} & F^{*} & H & I\\
D^{*} & G^{*} & I^{*} & J
\end{pmatrix}
\end{eqnarray}
where $A$,$B$,$C$,$D$,$E$,$F$,$G$,$H$,$I$,$J$ denote the $2\times 2$ block matrices.\\
When the partial transposition operation acts on the first qubit $A$ of the state $\rho_{ABC}$, the state transformed as
\begin{eqnarray}
\rho_{ABC}\rightarrow \rho_{ABC}^{T_{A}}\equiv [T\otimes I\otimes I](\rho_{ABC})
\label{PTA}
\end{eqnarray}
The partial transposition with respect to the second and third qubit respectively reduces the state $\rho_{ABC}$ to
\begin{eqnarray}
\rho_{ABC}\rightarrow \rho_{ABC}^{T_{B}}\equiv [I\otimes T\otimes I](\rho_{ABC})
\label{PTB}
\end{eqnarray}
\begin{eqnarray}
\rho_{ABC}\rightarrow \rho_{ABC}^{T_{C}}\equiv [I\otimes I\otimes T](\rho_{ABC})
\label{PTC}
\end{eqnarray}
The partial transposed states $\rho_{ABC}^{T_{A}}$, $\rho_{ABC}^{T_{B}}$, $\rho_{ABC}^{T_{C}}$ can be expressed in terms of block matrices as
\begin{eqnarray}
\rho_{ABC}^{T_{A}}=
\begin{pmatrix}
	A & B & C^{*} & F^{*}\\
	B^{*} & E & D^{*} & G^{*}\\
	C & D & H & I\\
	F & G & I^{*} & J
\end{pmatrix}
\label{PTA1}
\end{eqnarray}
\begin{eqnarray}
\rho_{ABC}^{T_{B}}=
\begin{pmatrix}
A & B^{*} & C & F\\
B & E & D & G\\
C^{*} & D^{*} & H & I^{*}\\
F^{*} & G^{*} & I & J
\end{pmatrix}
\label{PTB1}
\end{eqnarray}
\begin{eqnarray}
\rho_{ABC}^{T_{C}}=
\begin{pmatrix}
A^{*} & B^{*} & C^{*} & D^{*}\\
B & E^{*} & F^{*} & G^{*}\\
C & F & H^{*} & I^{*}\\
D & G & I & J^{*}
\end{pmatrix}
\label{PTC1}
\end{eqnarray}
It is well known that the partial transposition criterion is necessary and sufficient for $2 \otimes 2$ and $2 \otimes 3$ system while it is only necessary condition for the system $m \otimes n,~~(m\geq2,n\geq 3,~~\textrm{If}~~ m=2~~ \textrm{then}~~ n\neq 3)$ and for the multipartite system also. We now consider the simplest tripartite system i.e. $2 \otimes 2 \otimes 2$ quantum system to study its entanglement properties through partial transposition operation on any one of the single qubit of the given three-qubit system.\\
\textbf{(i)} Let us choose a particular form of an arbitrary state lying in $GHZ$-class, which is given by
\begin{eqnarray}
|GHZ\rangle_{ABC}=\alpha |000\rangle_{ABC}+ \beta|111\rangle_{ABC},~~|\alpha|^{2}+|\beta|^{2}=1
\label{ghz}
\end{eqnarray}
The density operator $\rho_{GHZ}=|GHZ\rangle_{ABC}\langle GHZ|$ can be expressed as
\begin{eqnarray}
\rho_{GHZ}=
\begin{pmatrix}
A_{1} & B_{1} & C_{1} & D_{1}\\
B_{1}^{*} & E_{1} & F_{1} & G_{1}\\
C_{1}^{*} & F_{1}^{*} & H_{1} & I_{1}\\
D_{1}^{*} & G_{1}^{*} & I_{1}^{*} & J_{1}
\end{pmatrix}
\end{eqnarray}
where $A_{1}=\begin{pmatrix}
|\alpha|^{2} & 0\\
0 & 0
\end{pmatrix}, D_{1}=\begin{pmatrix}
0 & \alpha\beta^{*}\\
0 & 0
\end{pmatrix},J_{1}=\begin{pmatrix}
0 & 0 \\
0 & |\beta|^{2}
\end{pmatrix}$ and all other $2 \times 2$ block matrices are null matrices.\\
If we apply partial transposition operation on the qubit $A$ of the state described by the density operator $\rho_{GHZ}$ then the partial transposed state $\rho^{T_{A}}_{GHZ}$ at the output can be obtained by the prescription given in (\ref{PTA1}). The eigenvalues of $\rho^{T_{A}}_{GHZ}$ are given by $\{0,0,0,0,|\alpha|^{2},|\beta|^{2},|\alpha||\beta|,-|\alpha||\beta|\}$. Thus, $\rho^{T_{A}}_{GHZ}$ has one negative eigenvalue. The minimum eigenvalue of $\rho^{T_{A}}_{GHZ}$ is given by
\begin{eqnarray}
\lambda_{min}(\alpha,\beta)=-|\alpha||\beta|
\label{mineigenvalue}
\end{eqnarray}
Since the minimum eigenvalue of $\rho^{T_{A}}_{GHZ}$ is negative so the state $\rho_{GHZ}$ under investigation is an entangled state for all non-zero values of the state parameter $\alpha$ and $\beta$. The most negative eigenvalue is important for more than one reason, which will be clear in the later stage.\\
The most negative eigenvalue can be obtained for $|\alpha|=\frac{1}{\sqrt{2}}$ and $|\beta|=\frac{1}{\sqrt{2}}$. Therefore, the minimum most eigenvalue of $\rho^{T_{A}}_{GHZ}$ is given by
\begin{eqnarray}
\lambda_{min}^{\rho^{T_{A}}_{GHZ}}=-\frac{1}{2}
\label{minmostAeigenvalue}
\end{eqnarray}
For the same value of $\alpha$ and $\beta$ i.e. for $|\alpha|=\frac{1}{\sqrt{2}}$ and $|\beta|=\frac{1}{\sqrt{2}}$, we can obtain the maximum value of tangle $\tau$ which is given by $\tau=1$.\\
Proceeding in a similar way, we can obtain the minimum most eigenvalue of $\rho^{T_{B}}_{GHZ}$ and $\rho^{T_{C}}_{GHZ}$ and they are given by
\begin{eqnarray}
\lambda_{min}^{\rho^{T_{B}}_{GHZ}}=\lambda_{min}^{\rho^{T_{C}}_{GHZ}}=-\frac{1}{2}
\label{minmostBCeigenvalue}
\end{eqnarray}
Since the minimum most eigenvalue of the partial transposed state with respect to the qubits $A$, $B$ and $C$ are same so we denote it by $\lambda_{min}^{\rho_{GHZ}}$. Thus, we have $\lambda_{min}^{\rho^{T_{A}}_{GHZ}}=\lambda_{min}^{\rho^{T_{B}}_{GHZ}}=\lambda_{min}^{\rho^{T_{C}}_{GHZ}}\equiv\lambda_{min}^{\rho_{GHZ}}$.\\
\textbf{(ii)} Next, we will choose a particular form of an arbitrary state belong to $W$-class, which is given by
\begin{eqnarray}
|W\rangle_{ABC}&=&\lambda_{0} |001\rangle_{ABC}+ \lambda_{1} |010\rangle_{ABC}+ \lambda_{2} |100\rangle_{ABC}
\label{wstate}
\end{eqnarray}
where the state parameters $\lambda_{i} (i=0,1,2)$ are real numbers satisfying $\lambda_{0}^{2}+\lambda_{1}^{2}+\lambda_{2}^{2}=1$. One may choose $W$-class states with more terms also for detailed analysis but may face difficulty in finding the analytical form of eigenvalues.\\
The density operator $\rho_{W}$ can be expressed as
\begin{eqnarray}
\rho_W=
\begin{pmatrix}
A & B & C & D\\
B^{*} & E & F & G\\
C^{*} & F^{*} & H & I\\
D^{*} & G^{*} & I^{*} & J\\
\end{pmatrix}
\end{eqnarray}
where,
\begin{eqnarray}
 A=
\begin{pmatrix}
0 & 0\\
0 & \lambda_0^{2}\\
\end{pmatrix},
B=
\begin{pmatrix}
0 & 0\\
\lambda_0\lambda_1 & 0\\
\end{pmatrix},
C=
\begin{pmatrix}
0 & 0\\
\lambda_0\lambda_2 & 0\\
\end{pmatrix},\nonumber\\
E=
\begin{pmatrix}
\lambda_1^{2} & 0\\
0 & 0\\
\end{pmatrix},
F=
\begin{pmatrix}
\lambda_1\lambda_2 & 0\\
0 & 0\\
\end{pmatrix},
H=
\begin{pmatrix}
\lambda_2^{2} & 0\\
0 & 0\\
\end{pmatrix},
\end{eqnarray}
D, G, I, J are zero matrices.\\
The eigenvalues of $\rho^{T_{A}}_{W}$ can be calculated as $\{0,0,0,0,\lambda_{0}^{2}+\lambda_{1}^{2},\lambda_{2}^{2},\lambda_{2}\sqrt{\lambda_{0}^{2}+\lambda_{1}^{2}},
-\lambda_{2}\sqrt{\lambda_{0}^{2}+\lambda_{1}^{2}}\}$. Thus, $\rho^{T_{A}}_{W}$ has one negative eigenvalue irrespective of the sign of the real parameters $\lambda_{0}$, $\lambda_{1}$, $\lambda_{2}$. The minimum eigenvalue of $\rho^{T_{A}}_{W}$ is given by
\begin{eqnarray}
\lambda_{min}(\lambda_{0},\lambda_{1},\lambda_{2})&=&-|\lambda_{2}|\sqrt{\lambda_{0}^{2}+\lambda_{1}^{2}}\nonumber\\&=&
-|\lambda_{2}|\sqrt{1-\lambda_{2}^{2}}
\label{mineigenvalueW}
\end{eqnarray}
In this case also, we find that the minimum eigenvalue of $\rho^{T_{A}}_{W}$ is negative so the state described by the density operator $\rho_{W}$ is an entangled state for all non-zero values of the state parameter $\lambda_{i} (i=0,1,2)$.\\
The most negative eigenvalue can be obtained for $|\lambda_{2}|=\frac{7}{10}$ and for any value of $\lambda_{0}$ and $\lambda_{1}$ satisfying $\lambda_{0}^{2}+\lambda_{1}^{2}=0.51$. Thus, the minimum most eigenvalue of $\rho^{T_{A}}_{GHZ}$ is given by
\begin{eqnarray}
\lambda_{min}^{\rho^{T_{A}}_{GHZ}}=-0.4999
\label{minmostAeigenvaluew}
\end{eqnarray}
Proceeding in a similar way, the eigenvalues of $\rho_W^{T_B}$ and $\rho_W^{T_C}$ can be calculated as $\{0,0,0,0,\lambda_1^{2}, \lambda_0^{2}+\lambda_2^{2}, \lambda_1\sqrt{\lambda_0^{2}+\lambda_2^{2}}, -\lambda_1\sqrt{\lambda_0^{2}+\lambda_2^{2}}
\}$ and $\{0,0,0,0,\lambda_0^{2}, \lambda_1^{2}+\lambda_2^{2}, \lambda_0\sqrt{\lambda_1^{2}+\lambda_2^{2}}, -\lambda_0\sqrt{\lambda_1^{2}+\lambda_2^{2}}\}$ respectively. Following the same procedure, we can obtain the minimum most eigenvalue of $\rho^{T_{B}}_{W}$ and $\rho^{T_{C}}_{W}$ respectively as
\begin{eqnarray}
\lambda_{min}^{\rho^{T_{B}}_{W}}=\lambda_{min}^{\rho^{T_{C}}_{W}}=-0.4999
\label{minmostBCeigenvaluew}
\end{eqnarray}
Since the minimum most eigenvalue of the partial transposed state with respect to the qubits $A$, $B$ and $C$ are same so we can denote it by $\lambda_{min}^{\rho_{W}}$. Thus, we have $\lambda_{min}^{\rho^{T_{A}}_{W}}=\lambda_{min}^{\rho^{T_{B}}_{W}}=\lambda_{min}^{\rho^{T_{C}}_{W}}=\lambda_{min}^{\rho_{W}}$.

\section{Structural physical approximation of partial transposition (SPA-PT) of a single qubit in a three-qubit system}
Let us consider a map which may be defined as the convex combination of the depolarizing map and the partial transposition map. We can mix the depolarizing map to the partial transposition map in such a way that the resulting map is completely positive. The newly constructed map can be considered as the approximation of partial transposition map and termed as the structural physical approximation (SPA) of the partial transposition (PT) with respect to the qubit $A$. It is denoted by $\widetilde{[T\otimes I\otimes I]}$. If we apply SPA-PT map on the qubit $A$ of the three-qubit state $\rho_{ABC}$ then the state transformed as
\begin{eqnarray}
\widetilde{[T\otimes I\otimes I]}\rho_{ABC}\equiv \widetilde{\rho^{T_{A}}}&=&\frac{p_{A}}{8}(I\otimes I\otimes I)+(1-p_{A})\times \nonumber\\&&[T\otimes I\otimes I](\rho_{ABC})
\label{SPA-PT1}
\end{eqnarray}
where $0\leq p_{A} \leq 1$.\\
In a similar way, SPA-PT with respect to the qubit $B$ and $C$ respectively transformed the state $\rho_{ABC}$ as
\begin{eqnarray}
\widetilde{[I\otimes T\otimes I]}\rho_{ABC}\equiv \widetilde{\rho^{T_{B}}}&=&\frac{p_{B}}{8}(I\otimes I\otimes I)+(1-p_{B})\times \nonumber\\&&[I\otimes T\otimes I](\rho_{ABC})
\label{SPA-PT2}
\end{eqnarray}
\begin{eqnarray}
\widetilde{[I\otimes I\otimes T]}\rho_{ABC}\equiv \widetilde{\rho^{T_{C}}}&=&\frac{p_{C}}{8}(I\otimes I\otimes I)+(1-p_{C})\times \nonumber\\&&[I\otimes I\otimes T]\rho_{ABC}
\label{SPA-PT3}
\end{eqnarray}
where $0\leq p_{B},p_{C} \leq 1$.\\
\subsection{When Structural Physical Approximation Map will be Completely Positive?}
Here we derive the condition for which the SPA-PT map is completely positive. We deduce the condition by considering the approximation of partial transposition operation with respect to the qubit $A$. In a similar fashion, one can deduce the same condition by approximating the partial transposition with respect to the other two qubits $B$ and $C$ respectively. \\
The SPA-PT map with respect to the qubit $A$ is positive if $\lambda_{min}(\widetilde{\rho^{T_{A}}})\geq 0$ holds. Therefore, using (\ref{SPA-PT1}), we can write the expression of minimum eigenvalue of the operator $\widetilde{\rho^{T_{A}}}$ as
\begin{eqnarray}
\lambda_{min}(\widetilde{\rho^{T_{A}}})&=&\lambda_{min}[\frac{p_{A}}{8}(I\otimes I\otimes I)+(1-p_{A})\nonumber\\&& \times(T\otimes I\otimes I)\rho_{ABC}]
\label{mineigvalA}
\end{eqnarray}
Further, the R.H.S of (\ref{mineigvalA}) can be reduced using Weyl's inequality as
\begin{eqnarray}
\lambda_{min}(\widetilde{\rho^{T_{A}}}) &\geq& \frac{p_{A}}{8}\lambda_{min}(I\otimes I\otimes I)+(1-p_{A})\nonumber\\&& \times \lambda_{min}[(T\otimes I\otimes I)\rho_{ABC}]
\label{ineq1}
\end{eqnarray}
If $\lambda_{min}[(T\otimes I\otimes I)\rho_{ABC}]\equiv\lambda_{min}(\rho^{T_{A}})\geq 0$, then the above inequality (\ref{ineq1}) reduces to
\begin{eqnarray}
\lambda_{min}(\widetilde{\rho^{T_{A}}}) &\geq& \frac{p_{A}}{8}
\label{SPA2}
\end{eqnarray}
Now, our task is to find out the minimum value of $p_{A}$ for which the operator
$\widetilde{(T\otimes I\otimes I)}$ will be completely positive. Since the partial transposition operator is not a completely positive operator so the induced map $[(I \otimes I \otimes I) \otimes (T \otimes I \otimes I)]$ generate at least one negative eigenvalues. The most negative eigenvalue generated when the induced map $[(I \otimes I \otimes I) \otimes (T \otimes I \otimes I)]$ is applying on the state $[(I \otimes I \otimes I) \otimes |GHZ\rangle_{ABC}]$, where $|GHZ\rangle_{ABC}=\frac{1}{\sqrt{2}}(|000\rangle+|111\rangle)$. Thus, if we suitably choose the minimum value of $p_{A}$ for which the positive eigenvalues of the maximally mixed three-qubit state generated by the depolarizing map dominate over the minimum most negative eigenvalue generated by the induced map then we can make the operator $\widetilde{(T\otimes I\otimes I)}$ completely positive. Therefore, the map $\widetilde{(T\otimes I\otimes I)}$ is completely positive and hence physically implementable when
\begin{eqnarray}
p_{A}\geq \frac{4}{5}
\label{cond1}
\end{eqnarray}
In a similar way, it can be shown that if we take the partial transposition with respect to system $B$ and $C$ then the SPA-PT map will be completely positive when
\begin{eqnarray}
p_{B}\geq \frac{4}{5}
\label{cond2}
\end{eqnarray}
\begin{eqnarray}
p_{C}\geq \frac{4}{5}
\label{cond3}
\end{eqnarray}
\subsection{Determination of the matrix elements of the density matrix after SPA-PT operation}
\noindent In this section, we study how the entries of the approximated partial transposed density matrix denoted by $\widetilde{\varrho_{ABC}}$ is related with the entries of the original matrix described by the density matrix $\varrho_{ABC}$. If we have an arbitrary three-qubit state described by the density operator $\varrho_{ABC}$ then after the application of SPA-PT operation on it, the density matrix has been changed and changes to $\widetilde{\varrho}_{ABC}$. As a consequence, the elements of the matrix $\widetilde{\varrho}_{ABC}$ can be expressed in terms of the matrix elements of $\varrho_{ABC}$. Thus, the determination of the matrix elements of the density matrix $\widetilde{\varrho}_{ABC}$ is important because the entanglement properties of $\varrho_{ABC}$ can be studied using the matrix elements of $\widetilde{\varrho}_{ABC}$.
To start with, we consider an arbitrary three-qubit quantum state described by the density matrix $\varrho_{ABC}$, which is given by
\begin{eqnarray}
\varrho_{ABC}=
\begin{pmatrix}
t_{11} & t_{12} & t_{13} & t_{14} & t_{15} & t_{16} & t_{17} & t_{18} \\
t_{12}^{*} & t_{22} & t_{23} & t_{24} & t_{25} & t_{26} & t_{27} & t_{28} \\
t_{13}^{*} & t_{23}^{*} & t_{33} & t_{34} & t_{35} & t_{36} & t_{37} & t_{38} \\
t_{14}^{*} & t_{24}^{*} & t_{34}^{*} & t_{44} & t_{45} & t_{46} & t_{47} & t_{48}\\
t_{15}^{*} & t_{25}^{*} & t_{35}^{*} & t_{45}^{*} & t_{55} & t_{56} & t_{57} & t_{58}\\
t_{16}^{*} & t_{26}^{*} & t_{36}^{*} & t_{46}^{*} & t_{56}^{*} & t_{66} & t_{67} & t_{68}\\
t_{17}^{*} & t_{27}^{*} & t_{37}^{*} & t_{47}^{*} & t_{57}^{*} & t_{67}^{*} & t_{77} & t_{78}\\
t_{18}^{*} & t_{28}^{*} & t_{38}^{*} & t_{48}^{*} & t_{58}^{*} & t_{68}^{*} & t_{78}^{*} & t_{88}
\end{pmatrix}, \sum_{i=1}^{8}t_{ii}=1
\label{threequbitstate}
\end{eqnarray}
where $(*)$ denotes the complex conjugate.\\\\
The SPA-PT with respect to qubit $A$ of a three-qubit quantum state $\varrho_{ABC}$ is given by
\begin{eqnarray}
\widetilde{\rho^{T_{A}}}=[\frac{1}{10}(I\otimes I\otimes I)+\frac{1}{5}(T\otimes I\otimes I)\rho_{ABC}]
\label{spaptthreequbit1}
\end{eqnarray}
where $T$ denotes the transposition operator acting on qubit $A$.\\
The matrix representation of $\widetilde{\rho^{T_{A}}}$ is given by
\begin{eqnarray}
\widetilde{\rho^{T_{A}}}=
\begin{pmatrix}
\tilde{t}_{11} & \tilde{t}_{12} & \tilde{t}_{13} & \tilde{t}_{14} & \tilde{t}_{15} & \tilde{t}_{16} & \tilde{t}_{17} & \tilde{t}_{18}\\
\tilde{t}_{12}^{*} & \tilde{t}_{22} & \tilde{t}_{23} & \tilde{t}_{24} & \tilde{t}_{25} & \tilde{t}_{26}  & \tilde{t}_{27} & \tilde{t}_{28} \\
\tilde{t}_{13}^{*} & \tilde{t}_{23}^{*} & \tilde{t}_{33} & \tilde{t}_{34} & \tilde{t}_{35} & \tilde{t}_{36}  & \tilde{t}_{37} & \tilde{t}_{38} \\
\tilde{t}_{14}^{*} & \tilde{t}_{24}^{*} & \tilde{t}_{34}^{*} & \tilde{t}_{44} & \tilde{t}_{45} & \tilde{t}_{46}  & \tilde{t}_{47} & \tilde{t}_{48}\\
\tilde{t}_{15}^{*} & \tilde{t}_{25}^{*} & \tilde{t}_{35}^{*} & \tilde{t}_{45}^{*} & \tilde{t}_{55} & \tilde{t}_{56}  & \tilde{t}_{57} & \tilde{t}_{58}\\
\tilde{t}_{16}^{*} & \tilde{t}_{26}^{*} & \tilde{t}_{36}^{*} & \tilde{t}_{46}^{*} & \tilde{t}_{56}^{*} & \tilde{t}_{66}  & \tilde{t}_{67} & \tilde{t}_{68}\\
\tilde{t}_{17}^{*} & \tilde{t}_{27}^{*} & \tilde{t}_{37}^{*} & \tilde{t}_{47}^{*} & \tilde{t}_{57}^{*} & \tilde{t}_{67}^{*}  & \tilde{t}_{77} & \tilde{t}_{78}\\
\tilde{t}_{18}^{*} & \tilde{t}_{28}^{*} & \tilde{t}_{38}^{*} & \tilde{t}_{48}^{*} & \tilde{t}_{58}^{*} & \tilde{t}_{68}^{*}  & \tilde{t}_{78}^{*} & \tilde{t}_{88}\\
\end{pmatrix}, \sum_{i=1}^{8}\tilde{t}_{ii}=1
\label{qutrit-qubit2}
\end{eqnarray}
where the entries of the density matrix $\widetilde{\rho^{T_{A}}}$ are given by,
\begin{eqnarray}
&&\tilde{t}_{11}=\frac{1}{10}+\frac{t_{11}}{5}, \tilde{t}_{12}=\frac{t_{12}}{5}, \tilde{t}_{13}=\frac{t_{13}}{5}, \tilde{t}_{14}=\frac{t_{14}}{5}\nonumber\\&& \tilde{t}_{15}=\frac{t_{15}^{*}}{5}, \tilde{t}_{16}=\frac{t_{25}^{*}}{5}, \tilde{t}_{17}=\frac{t_{35}^{*}}{5}, \tilde{t}_{18}=\frac{t_{45}^{*}}{5}\nonumber\\&& \tilde{t}_{22}=\frac{1}{10}+\frac{t_{22}}{5}, \tilde{t}_{23}=\frac{t_{23}}{5}, \tilde{t}_{24}=\frac{t_{24}}{5}, \tilde{t}_{25}=\frac{t_{16}^{*}}{5}\nonumber\\&& \tilde{t}_{26}=\frac{t_{26}^{*}}{5}, \tilde{t}_{27}=\frac{t_{36}^{*}}{5}, \tilde{t}_{28}=\frac{t_{46}^{*}}{5}, \tilde{t}_{33}=\frac{1}{10}+\frac{t_{33}}{5}\nonumber\\&& \tilde{t}_{34}=\frac{t_{34}}{5}, \tilde{t}_{35}=\frac{t_{17}^{*}}{5}, \tilde{t}_{36}=\frac{t_{27}^{*}}{5}, \tilde{t}_{37}=\frac{t_{37}^{*}}{5}\nonumber\\&& \tilde{t}_{38}=\frac{t_{47}^{*}}{5}, \tilde{t}_{44}=\frac{1}{10}+\frac{t_{44}}{5}, \tilde{t}_{45}=\frac{t_{18}^{*}}{5}, \tilde{t}_{46}=\frac{t_{28}^{*}}{5}\nonumber\\&& \tilde{t}_{47}=\frac{t_{38}^{*}}{5}, \tilde{t}_{48}=\frac{t_{48}^{*}}{5}, \tilde{t}_{55}=\frac{1}{10}+\frac{t_{55}}{5},  \tilde{t}_{56}=\frac{t_{56}}{5}\nonumber\\&& \tilde{t}_{57}=\frac{t_{57}}{5}, \tilde{t}_{58}=\frac{t_{58}}{5}, \tilde{t}_{66}=\frac{1}{10}+\frac{t_{66}}{5},  \tilde{t}_{67}=\frac{t_{67}}{5}\nonumber\\&& \tilde{t}_{68}=\frac{t_{68}}{5}, \tilde{t}_{77}=\frac{1}{10}+\frac{t_{77}}{5},  \tilde{t}_{78}=\frac{t_{78}}{5}\nonumber\\&& \tilde{t}_{88}=\frac{1}{10}+\frac{t_{88}}{5}
\label{spa1}
\end{eqnarray}
Following the same procedure, one can determine the matrix elements of the density matrix resulting from the application of completely positive maps $\widetilde{[I\otimes T\otimes I]}\rho_{ABC}$ and $\widetilde{[I\otimes I\otimes T]}\rho_{ABC}$ respectively.\\
In the next section, we will show that the minimum eigenvalue of $\widetilde{\rho^{T_{A}}}$, $\widetilde{\rho^{T_{B}}}$ and $\widetilde{\rho^{T_{C}}}$ is the entity that may detect whether the given three-qubit state $\varrho_{ABC}$ possesses the property of entanglement or not so it is very essential to extract the information about the entries of the matrix $\widetilde{\rho^{T_{A}}}$, $\widetilde{\rho^{T_{B}}}$ and $\widetilde{\rho^{T_{C}}}$. Thus the matrix elements given by (\ref{spa1}) plays a vital role in detecting the entanglement of three-qubit system when SPA-PT operation is performing with respect to the system $A$.
 
\section{Necessary condition for the separability (either in the form of a full separability or biseparability) of a three-qubit state}
In this section, we will derive the necessary condition for the full separability and biseparability of a three-qubit state. Thus, if any three-qubit state violating the necessary condition then we can infer that the given three-qubit state is a genuine entangled state.\\
To move forward in this direction, we consider any three-qubit state shared between three distant parties $Alice (A)$, $Bob (B)$ and $Charlie (C)$ and ask whether the shared state is entangled or not?
To detect the entanglement in three-qubit system, one may follow the partial transposition criterion and thus apply partial transposition operation on any one of the qubit of the given three-qubit system. To overcome the difficulty of the real implementation of partial transposition map in an experiment, we approximate the partial transposition operation by the method of structural physical approximation. We have already shown in the previous section that the SPA-PT map can serve as a completely positive map and thus can be implemented in a real experimental set up, if we add sufficient proportion of depolarizing map to the partial transposition map. Now we are in a position to give the statement of a necessary condition of the separability and biseparability of a three-qubit state.\\
\textbf{Theorem-1:} If the state described by the density operator $\rho_{ABC}$ denoting either a separable state of the form $A-B-C$ or a biseparable state of the form $A-BC$ then the following inequality is satisfied
\begin{eqnarray}
\lambda_{min}(\widetilde{\rho^{T_{A}}}) &\geq& \frac{1}{10}
\label{finalcond1}
\end{eqnarray}
\textbf{Proof:} The required inequality (\ref{finalcond1}) follows from (\ref{SPA2}) and (\ref{cond1}).\\
\textbf{Theorem-2:} If the state described by the density operator $\rho_{ABC}$ denoting either a separable state of the form $A-B-C$ or a biseparable state of the form $B-AC$ then the following inequality is satisfied
\begin{eqnarray}
\lambda_{min}(\widetilde{\rho^{T_{B}}}) &\geq& \frac{1}{10}
\label{finalcond2}
\end{eqnarray}
\textbf{Theorem-3:} If the state described by the density operator $\rho_{ABC}$ denoting either a separable state of the form $A-B-C$ or a biseparable state of the form $C-AB$ then the following inequality is satisfied
\begin{eqnarray}
\lambda_{min}(\widetilde{\rho^{T_{C}}}) &\geq& \frac{1}{10}
\label{finalcond3}
\end{eqnarray}
Let us now provide few results that may help in detecting the given three-qubit state $\rho_{ABC}$ as either a separable state, biseparable state or a genuine entangled state.
To do this task, we assume that $\lambda_{min}(\widetilde{\rho})=max\{\lambda_{min}(\widetilde{\rho^{T_{A}}}), \lambda_{min}(\widetilde{\rho^{T_{B}}}), \lambda_{min}(\widetilde{\rho^{T_{C}}})\}$.\\
\textbf{Result-1:} If $\lambda_{min}(\widetilde{\rho}) <\frac{1}{10}$, then $\rho_{ABC}$ is a genuine entangled state.\\
\textbf{Result-2:} If $\lambda_{min}(\widetilde{\rho^{T_{A}}})\geq\frac{1}{10}$ and either $\lambda_{min}(\widetilde{\rho^{T_{B}}})<\frac{1}{10}$ or  $\lambda_{min}(\widetilde{\rho^{T_{C}}})<\frac{1}{10}$ or both $\lambda_{min}(\widetilde{\rho^{T_{B}}}),\lambda_{min}(\widetilde{\rho^{T_{C}}})<\frac{1}{10}$ holds, then $\rho_{ABC}$ is biseparable in $A-BC$ cut.\\
\textbf{Result-3:} If $\lambda_{min}(\widetilde{\rho^{T_{B}}})\geq\frac{1}{10}$ and either $\lambda_{min}(\widetilde{\rho^{T_{A}}})<\frac{1}{10}$ or  $\lambda_{min}(\widetilde{\rho^{T_{C}}})<\frac{1}{10}$ or both $\lambda_{min}(\widetilde{\rho^{T_{A}}}),\lambda_{min}(\widetilde{\rho^{T_{C}}})<\frac{1}{10}$ holds, then $\rho_{ABC}$ is biseparable in $B-AC$ cut.\\
\textbf{Result-4:} If $\lambda_{min}(\widetilde{\rho^{T_{C}}})\geq\frac{1}{10}$ and either $\lambda_{min}(\widetilde{\rho^{T_{A}}})<\frac{1}{10}$ or  $\lambda_{min}(\widetilde{\rho^{T_{B}}})<\frac{1}{10}$ or both $\lambda_{min}(\widetilde{\rho^{T_{A}}}),\lambda_{min}(\widetilde{\rho^{T_{B}}})<\frac{1}{10}$ holds, then $\rho_{ABC}$ is biseparable in $C-AB$ cut.\\
\textbf{Result-5:} If $\lambda_{min}(\widetilde{\rho^{T_{A}}})\geq\frac{1}{10}$, $\lambda_{min}(\widetilde{\rho^{T_{B}}})\geq \frac{1}{10}$ and  $\lambda_{min}(\widetilde{\rho^{T_{C}}}) \geq \frac{1}{10}$ holds, then $\rho_{ABC}$ is a fully separable state.\\

\section{A Few Examples}
In this section, we discuss about few examples of three-qubit genuine entangled states and three-qubit biseparable states that can be detected by the results given in the previous section.
\subsection{Genuine Entangled States}
\textbf{Example-1:} Let us consider the state $|\psi_{G_{1}}\rangle$ described by the density operator $\rho_{G_{1}}=|\psi_{G_{1}}\rangle\langle \psi_{G_{1}}|$, where $|\psi_{G_{1}}\rangle =\alpha|000\rangle +\beta|111\rangle,~~ |\alpha|^2+|\beta|^2=1$. We now proceed to calculate the minimum eigenvalue of $\widetilde{\rho_{G_{1}}^{T_{A}}}$, $\widetilde{\rho_{G_{1}}^{T_{B}}}$ and $\widetilde{\rho_{G_{1}}^{T_{C}}}$.\\
The eigenvalues are given by
\begin{eqnarray}
\lambda_{min}(\widetilde{\rho_{G_{1}}^{T_{A}}})=\lambda_{min}(\widetilde{\rho_{G_{1}}^{T_{B}}})=\lambda_{min}(\widetilde{\rho_{G_{1}}^{T_{C}}})&\equiv&
\lambda_{min}(\widetilde{\rho_{G_{1}}})\nonumber\\&=&\frac{1-2\alpha\beta}{10}
\label{ex1}
\end{eqnarray}
Thus, we can easily find that $\lambda_{min}(\widetilde{\rho_{G_{1}}})<\frac{1}{10}$ when $\alpha\beta>0$. Applying \textbf{Result-1}, we can say that the state $|\psi_{G_{1}}\rangle$ represent genuine entangled state when $\alpha\beta>0$.\\

\noindent \textbf{Example-2:}  Let us consider a pure three-qubit state which is given by
\begin{eqnarray}
|\psi_{G_{2}}\rangle =\frac{1}{\sqrt{5}}[|000\rangle+|100\rangle+|101\rangle+|110\rangle+|111\rangle]
\label{ex2}
\end{eqnarray}
For the given state $|\psi_{G_{2}}\rangle$, we have
\begin{eqnarray}
&&\lambda_{min}(\widetilde{|\psi_{G_{2}}\rangle^{T_{A}}\langle\psi_{G_{2}}|})=0.030718,
\lambda_{min}(\widetilde{|\psi_{G_{2}}\rangle^{T_{B}}\langle\psi_{G_{2}}|})=\nonumber\\&&
\lambda_{min}(\widetilde{|\psi_{G_{2}}\rangle^{T_{C}}\langle\psi_{G_{2}}|})=0.0434315
\label{ex21}
\end{eqnarray}
Therefore, we find that $\lambda_{min}(\widetilde{|\psi_{G_{2}}\rangle\langle\psi_{G_{2}}|})=max\{0.030718,0.0434315\}=0.0434315$. Thus, $\lambda_{min}(\widetilde{|\psi_{G_{2}}\rangle\langle\psi_{G_{2}}|})<\frac{1}{10}$. Hence, the given state $|\psi_{G_{2}}\rangle\langle\psi_{G_{2}}|$ is a genuine entangled state.\\\\
\textbf{Example-3:}  Let us take another state defined by $\rho_{G_{3}}=|\psi_{G_{3}}\rangle\langle \psi_{G_{3}}|$, where $|\psi_{G_{3}}\rangle =\lambda_0|000\rangle+\lambda_1|100\rangle +\lambda_2|111\rangle,0 \leq \lambda_i \leq 1, \sum_{i=0}^{2}\lambda_i^{2}=1$.
For the given state $\rho_{G_{3}}$, we can easily verify that for all values of $\lambda_0$, $\lambda_1$ and $\lambda_2$ lying between 0 and 1, we have, $\lambda_{min}(\widetilde{\rho_{G_{3}}})<\frac{1}{10}$. Further, we have calculated the values of $\lambda_{min}(\widetilde{\rho_{G_{3}}})$ by taking some values of $\lambda_0$, $\lambda_1$ and $\lambda_2$ and those values are tabulated in the Table-I for the verification of our result. Thus from Result-1, the given state $\rho_{G_{3}}$ is a genuine three-qubit entangled state.\\\\
\textbf{Example-4}  Consider the state defined by $\rho_{GHZ,W}=q|GHZ\rangle \langle GHZ|+(1-q)|W\rangle \langle W|, ~~0 \leq q \leq 1$, where $|GHZ\rangle=\frac{1}{\sqrt{2}}[|000\rangle+|111\rangle]$ and $|W\rangle =\frac{1}{\sqrt{3}}[|001\rangle+|010\rangle+|100\rangle]$. For the given state described by the density operator $\rho_{GHZ,W}$, the minimum eigenvalues are given by
\begin{eqnarray}
\lambda_{min}(\widetilde{\rho_{GHZ,W}^{T_{A}}})&=&\lambda_{min}(\widetilde{\rho_{GHZ,W}^{T_{B}}})=\lambda_{min}(\widetilde{\rho_{GHZ,W}^{T_{C}}})
\nonumber\\&=&min\{Q_{1},Q_{2}\}
\end{eqnarray}
where $Q_{1}=\frac{1}{30}(4-q-\sqrt{1-2q+10q^2})$, and $Q_{2}=\frac{1}{60}(6+3q-\sqrt{32-64q+41q^2})$. It can be easily seen that $min\{Q_{1},Q_{2}\}<\frac{1}{10}$ and thus, we have $\lambda_{min}(\widetilde{rho_{GHZ,W}})<\frac{1}{10}$. Hence from \textbf{Result-1}, We can say that the given state $\rho_{GHZ,W}$ is a genuine entangled state for all $q \in [0,1]$.
\begin{table}
\begin{center}
\caption{Table varifying Result-I(i) for differnt values of the state parameters}
\begin{tabular}{|c|c|c|c|}\hline
State parameter & Minimum  & Minimum  & $\lambda_{min}(\widetilde{\rho_{G_{3}}})$\\ $(\lambda_0, \lambda_1, \lambda_2)$ & eigenvalue of &  eigenvalue of   & \\ & SPA-PT state w.r.t  & SPA-PT state w.r.t  & \\ & qubit $A$ and $C$ &  qubit $B$   & \\ & $\lambda_{min}(\widetilde{\rho_{G_{3}}^{T_{A}}})$,$\lambda_{min}(\widetilde{\rho_{G_{3}}^{T_{C}}})$ & $\lambda_{min}(\widetilde{\rho_{G_{3}}^{T_{B}}})$& \\   \hline
(0.7, 0.1, 0.707107)  & 0.00101 & $1.295\times 10^{-18}$ & 0.00101\\\hline
(0.3,0.4,0.866)  & 0.048 & 0.0134 & 0.048\\\hline
(0.7,0.3, 0.648) & 0.0093 & 0.0013 & 0.0093\\\hline
(0.1, 0.2, 0.9747) & 0.0805 & 0.056 & 0.0805\\\hline
(0.2, 0.4, 0.8944) & 0.0642 & 0.02 &  0.0642\\\hline
\end{tabular}
\end{center}
\end{table}\\
\subsection{Biseparable states}
\textbf{Example-1:}  Consider the state defined by the density matrix $\rho_{B_{1}}=q|0\rangle\langle 0| \otimes |\phi^{+}\rangle \langle \phi^{+}|+(1-q)|1\rangle \langle 1|\otimes |\phi^{-}\rangle \langle \phi^{-}|, ~~0 \leq q \leq 1$, where $|\phi^{\pm}\rangle=\frac{1}{\sqrt{2}}[|00\rangle \pm |11\rangle]$. For the given state $\rho_{B_{1}}$, the minimum eigenvalues of the partial transposed state are given by $\lambda_{min}(\widetilde{\rho_{B_{1}}^{T_{A}}})=\frac{1}{10}, \lambda_{min}(\widetilde{\rho_{B_{1}}^{T_{B}}})=\lambda_{min}(\widetilde{\rho_{B_{1}}^{T_{C}}})=\min\{\frac{q}{10}, \frac{1-q}{10}\}$. When the state parameter $q$ satisfying the inequality $0 \leq q \leq 1$, we observe that the minimum eigenvalue satisfy
\begin{eqnarray}
\lambda_{min}(\widetilde{\rho_{B_{1}}^{T_{A}}})=\frac{1}{10},~~\lambda_{min}(\widetilde{\rho_{B_{1}}^{T_{B}}})=
\lambda_{min}(\widetilde{\rho_{B_{1}}^{T_{C}}})<\frac{1}{10}
\end{eqnarray}
Therefore, we can infer from $\textbf{Result-2}$ that the given state $\rho_{B_{1}}$ is biseparable in $A-BC$ cut.\\

\textbf{Example-2}  Let us take a pure state, which is defined by $\rho_{B_{2}}=|\psi\rangle_{B_{2}}\langle \psi|$, where $|\psi\rangle_{B_{2}} =\lambda_0|001\rangle+\lambda_1|101\rangle +\lambda_2|111\rangle,~~\sum_{i=0}^{2}\lambda_{i}^{2}=1,~~0 \leq \lambda_{i} \leq 1, (i=0,1,2)$.
We find that for the given state described by the density operator $\rho_{B_{2}}$ that for all values of $\lambda_0, \lambda_1,\lambda_2 \in [0,1]$, we have, $\lambda_{min}(\widetilde{\rho_{B_{2}}^{T_{A}}})<\frac{1}{10}$, $\lambda_{min}(\widetilde{\rho_{B_{2}}^{T_{B}}})<\frac{1}{10}$ and $\lambda_{min}(\widetilde{\rho_{B_{2}}^{T_{C}}})\geq \frac{1}{10}$ . We have constructed $Table-II$ to clarify our result. Thus using $Result-4$, we can conclude that the given state $\rho_{B_{2}}$ is biseparable in $AB-C$ cut.\\
\begin{table}
\begin{center}
\caption{Table varifying Result-I(4) for differnt values of the state parameters}
\begin{tabular}{|c|c|c|c|}\hline
State parameter & Minimum  & Minimum  & $\lambda_{min}(\widetilde{\rho_{B_{2}}})$\\ $(\lambda_0, \lambda_1, \lambda_2)$ & eigenvalue of &  eigenvalue of   & \\ & SPA-PT state w.r.t  & SPA-PT state w.r.t  & \\ & qubit $A$ and $B$ &  qubit $C$   & \\ & $\lambda_{min}(\widetilde{\rho_{B_{2}}^{T_{A}}})$,$\lambda_{min}(\widetilde{\rho_{B_{2}}^{T_{B}}})$ & $\lambda_{min}(\widetilde{\rho_{B_{2}}^{T_{C}}})$& \\ \hline
(0.1, 0.4, 0.911)  & 0.0818 & 0.1 & 0.1\\\hline
(0.2, 0.4, 0.8944)  & 0.0642 & 0.1 & 0.1\\\hline
(0.6, 0.1, 0.7937) & 0.00475 & 0.1 & 0.1\\\hline
(0.5, 0.4, 0.7681) & 0.0232 & 0.1 & 0.1\\\hline
\end{tabular}
\end{center}
\end{table}
\subsection{Separable States}
\textbf{Example-1:} Let us consider the state also known as Kye state, which is defined by
\begin{eqnarray}
\rho_{S_{1}}=\frac{1}{8+8a}
\begin{pmatrix}
4+a & 0 & 0 & 0 & 0 & 0 & 0 & 2 \\
0 & a & 0 & 0 & 0 & 0 & 2 & 0 \\
0 & 0 & a & 0 & 0 & -2 & 0 & 0 \\
0 & 0 & 0 & a & 2 & 0 & 0 & 0\\
0 & 0 & 0 & 2 & a & 0 & 0 & 0\\
0 & 0 & -2 & 0 & 0 & a & 0 & 0\\
0 & 2 & 0 & 0 & 0 & 0 & a & 0\\
2 & 0 & 0 & 0 & 0 & 0 & 0 & 4+a
\end{pmatrix}
\end{eqnarray}
For the given state $\rho_{S_{1}}$, we find that $\lambda_{min}(\widetilde{\rho_{S_{1}}^{T_{A}}})=\lambda_{min}(\widetilde{\rho_{S_{1}}^{T_{B}}})=\lambda_{min}(\widetilde{\rho_{S_{1}}^{T_{C}}})=
\frac{2+5a}{40(1+a)}$. Thus, $\lambda_{min}(\widetilde{\rho_{S_{1}}})=\frac{2+5a}{40(1+a)}$. It has been found that for $a\geq 4$, $\frac{2+5a}{40(1+a)}>\frac{1}{10}$. Hence from \textbf{Result-5}, we conclude that the given state $\rho_{S_{1}}$ is fully separable for $a\geq 4$.\\\\
\textbf{Example-2:}  Consider another state defined by $\rho_{S_{2}}=(1-\alpha)|GHZ\rangle \langle GHZ|+\frac{\alpha}{8}I_8, ~~0 \leq \alpha \leq 1$, where $|GHZ\rangle=\frac{1}{\sqrt{2}}[|000\rangle+|111\rangle]$. For the given state $\rho_{S_{2}}$, the minimum eigenvalues of $\rho_{S_{2}}^{T_{A}}$, $\rho_{S_{2}}^{T_{B}}$ and $\rho_{S_{2}}^{T_{C}}$ are given by $\lambda_{min}(\widetilde{\rho_{S_{2}}^{T_{A}}})=\lambda_{min}(\widetilde{\rho_{S_{2}}^{T_{B}}})=\lambda_{min}(\widetilde{\rho_{S_{2}}^{T_{C}}})=
\frac{\alpha+4}{40}$. Thus,  $\lambda_{min}(\widetilde{\rho_{S_{2}}})=\frac{\alpha+4}{40}$. For $0.8 <\alpha  \leq 1$, we can check that $\lambda_{min}(\widetilde{\rho_{S_{2}}})>0.1$. From \textbf{Result-5}, we can say that the given state $\rho_{S_{2}}$ is fully separable for the state parameter $\alpha$ satisfying $0.8 <\alpha  \leq 1$.\\\\
\textbf{Example-3:}  Let us consider the state described by the density operator $\rho_{S_{3}}=q|\psi\rangle \langle \psi|+(1-q)|111\rangle \langle 111|, ~~0 \leq q \leq 1$, where $|\psi\rangle=\frac{1}{\sqrt{2}}[|001\rangle+|101\rangle]$. The minimum eigenvalues of the partial transposed states are given by $\lambda_{min}(\widetilde{\rho_{S_{3}}^{T_{A}}})=\lambda_{min}(\widetilde{\rho_{S_{3}}^{T_{B}}})=\lambda_{min}(\widetilde{\rho_{S_{3}}^{T_{C}}})=
\frac{1}{10}$. Thus, $\lambda_{min}(\widetilde{\rho_{S_{3}}})= \frac{1}{10}$. Therefore, \textbf{Result-5} tells us that $\rho_{S_{3}}$ is a fully separable state.\\\\
\subsection{Genuine/Biseparable/Separable}
\textbf{Example-1}  Let us consider the state defined by $\rho_{1}=q|000\rangle \langle 000|+(1-q)|GHZ\rangle \langle GHZ|, ~~0 \leq q \leq 1$, where $|GHZ\rangle=\frac{1}{\sqrt{2}}[|000\rangle+|111\rangle]$. For the given state $\rho_{1}$, we find that $\lambda_{min}(\widetilde{\rho_{1}^{T_{A}}})=\lambda_{min}(\widetilde{\rho_{1}^{T_{B}}})=\lambda_{min}(\widetilde{\rho_{1}^{T_{C}}})=\frac{q}{10}$. Thus,
\begin{eqnarray}
\lambda_{min}(\widetilde{\rho_{1}})&=&\frac{q}{10},~~~ 0 \leq q <1 \nonumber\\
&=&\frac{1}{10},~ q=1
\end{eqnarray}
Hence, $\rho_{1}$ is a genuine entangled state for $0\leq q<1$ and fully separable state for $q=1$ .\\
\textbf{Example-2:} Let us consider a mixed state, which is a convex combination of $GHZ$, $W$ and $\tilde{W}$ state and it is defined as\cite{jung}
\begin{eqnarray}
\rho_{2}&=&q_1 |GHZ\rangle \langle GHZ|+q_2|W\rangle \langle W|\nonumber\\&+&(1-q_1-q_2)|\tilde{W} \rangle \langle \tilde{W}|, 0\leq q_{1},q_{2}\leq 1
\end{eqnarray}
where,
\begin{eqnarray}
|GHZ\rangle =\frac{1}{\sqrt{2}}[|000\rangle +|111\rangle]\nonumber\\
|W\rangle=\frac{1}{\sqrt{3}}[|001\rangle+|010\rangle +|100\rangle ]\nonumber\\
|\tilde{W}\rangle=\frac{1}{\sqrt{3}}[|110\rangle +|101\rangle +|011\rangle]
\end{eqnarray}
The minimum eigenvalue of SPA-PT of the state $\rho_{2}$ is given by $\lambda_{min}(\widetilde{\rho_{2}^{T_{A}}})=\lambda_{min}(\widetilde{\rho_{2}^{T_{B}}})=\lambda_{min}(\widetilde{\rho_{2}^{T_{C}}})
=\frac{1}{30}(4-q_1-\sqrt{1-2q_1+10q_1^2-4q_2+4q_1q_2+4q_2^{2}})$. If the state parameters $q_{1}$ lying in the range $0.25 \leq q_1 \leq 1$ and $q_2=\frac{1-q_1}{n}$, where $n$ denote a positive integer, then within this range of parameters, the minimum eigenvalues of $\widetilde{\rho_{2}^{T_{A}}}$, $\widetilde{\rho_{2}^{T_{B}}}$ and $\widetilde{\rho_{2}^{T_{C}}}$ satisfying the inequality given by $\lambda_{min}(\widetilde{\rho_{2}^{T_{A}}})=\lambda_{min}(\widetilde{\rho_{2}^{T_{B}}})=\lambda_{min}(\widetilde{\rho_{2}^{T_{C}}})< \frac{1}{10}$. Thus, applying \textbf{Result-1}, we find that $\rho_{1}$ is a genuine entangled state.\\
Now, our task is to classify two inequivalent classes of genuine entangled state by considering different cases.\\
\textbf{Case-I:} If the state parameters $q_{1}$ lying in the range $0.25 \leq q_1 \leq 0.6269$ and $q_2=\frac{1-q_1}{n}$ then it has been shown that the three tangle of $\rho_{2}$ is zero \cite{jung}. Therefore, we find a sub-region in which not only three tangle of $\rho_{2}$ vanishes but also the state $\rho_{2}$ is genuine entangled. Hence, we can conclude that the state described by the density operator $\rho_{2}$ represent a $W$ class of state when the state parameters $q_{1}$ and $q_{2}$ satisfying $0.25 \leq q_1 \leq 0.629,~~q_2=\frac{1-q_1}{n}$.\\
\textbf{Case-II:} If the state parameters $q_{1}$ lying in the range $0.6269 \leq q_1 \leq 1$ and $q_2=\frac{1-q_1}{n}$ then the three tangle of genuine entangled state $\rho_{2}$ is non-zero. Therefore, the state $\rho_{2}$ represent a $GHZ$ class of state when the state parameters $q_{1}$ and $q_{2}$ satisfying $0.6269 < q_1 \leq 1,~~q_2=\frac{1-q_1}{n}$.

\section{Conclusion}
To summarize, we have studied the effect of partial transposition operation on one qubit of a three-qubit system. We have provided a matrix representation of three-qubit partially transposed states, in terms of $2\times 2$ block matrices, when partial transposition operation is performed with respect to first qubit or the second qubit or the third qubit. Then, we have studied how SPA map can be performed on a three-qubit system and explicitly claculated the matix elements of the matrix corresponding to the SPA-PT of a three-qubit state. Later, we have proposed different criterion for the classification of all possible SLOCC inequivalent classes of pure as well as mixed three qubit states. Our criterion is based on the method of SPA-PT map, which makes it experimentally realizable. Thus, using our experimental-friendly criterion, one can classify all possible SLOCC inequivalent classes in a three-qubit system. In the last section, we have supported our work with few examples.

\section{Acknowledgement}
A.K. would like to acknowledge the financial support from CSIR. This work is supported by CSIR File No. 08/133(0027)/2018-EMR-1.

\end{document}